\documentclass[letterpaper]{article} 
\usepackage{aaai2026}  
\usepackage{times}  
\usepackage{helvet}  
\usepackage{courier}  
\usepackage[hyphens]{url}  
\usepackage{graphicx} 
\usepackage{natbib}  
\usepackage{caption} 
\usepackage{algorithm}
\usepackage{amsmath}
\usepackage{algorithmic}
\usepackage{multirow}

\usepackage{newfloat}
\usepackage{listings}
\DeclareCaptionStyle{ruled}{labelfont=normalfont,labelsep=colon,strut=off} 
\floatstyle{ruled}
\newfloat{listing}{tb}{lst}{}
\floatname{listing}{Listing}
\title{Retrieval Feedback Memory Enhancement Large Model \\ Retrieval Generation Method}
\author{
    Leqian Li\textsuperscript{\rm 1},
    Dianxi Shi\textsuperscript{\rm 2}\thanks{Corresponding author.},
    Jialu Zhou\textsuperscript{\rm 1},
    Xinyu Wei\textsuperscript{\rm 1},
    Mingyue Yang\textsuperscript{\rm 1},
    Songchang Jin\textsuperscript{\rm 3},
    Shaowu Yang\textsuperscript{\rm 1}
}
\affiliations{
    \textsuperscript{\rm 1}College of Computer Science and Technology, National University of Defense Technology, china\\
    \textsuperscript{\rm 2}Advanced Institute of Big Data, Beijing, China\\
    \textsuperscript{\rm 3}Intelligent Game and Decision Lab, Beijing 100091, China

    \{llq,dxshi,zhoujialu23,xinyuwei,yangmingyue216,shaowu.yang\}@nudt.edu.cn
}

\usepackage{bibentry}

\begin{document}

\maketitle

\begin{abstract}
Large Language Models (LLMs) have shown remarkable capabilities across diverse tasks, yet they face inherent limitations such as constrained parametric knowledge and high retraining costs. Retrieval-Augmented Generation (RAG) augments the generation process by retrieving externally stored knowledge absent from the model’s internal parameters. However, RAG methods face challenges such as information loss and redundant retrievals during multi-round queries, accompanying the difficulties in precisely characterizing knowledge gaps for complex tasks. To address these problems, we propose Retrieval Feedback and Memory Retrieval Augmented Generation(RFM-RAG), which transforms the stateless retrieval of previous methods into stateful continuous knowledge management by constructing a dynamic evidence pool. Specifically, our method generates refined queries describing the model’s knowledge gaps using relational triples from questions and evidence from the dynamic evidence pool; Retrieves critical external knowledge to iteratively update this evidence pool; Employs a R-Feedback Model to evaluate evidence completeness until convergence. Compared to traditional RAG methods, our approach enables persistent storage of retrieved passages and effectively distills key information from passages to construct clearly new queries. Experiments on three public QA benchmarks demonstrate that RFM-RAG outperforms previous methods and improves overall system accuracy.
\end{abstract}

\section{Introduction}
\begin{figure}[t!]
	\centering
    \includegraphics[scale=0.82]{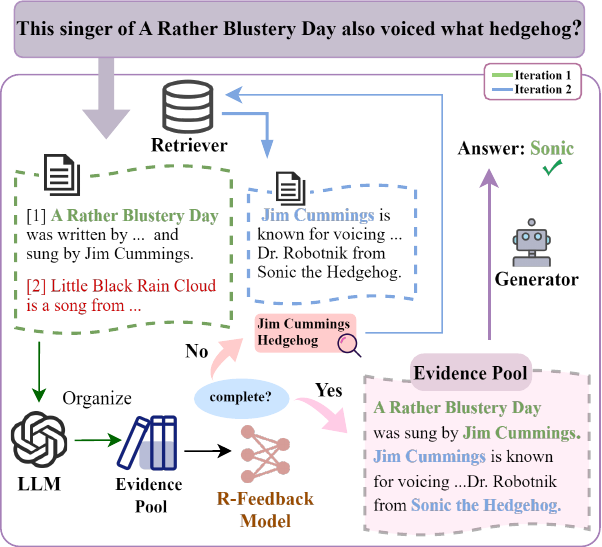}
	\caption{RFM-RAG employs an LLM to distill knowledge from retrieved results, dynamically updating an evidence pool. A R-Feedback Model then assesses the pool's completeness. If sufficient, evidence is passed to the generation model for final response. Otherwise, we formulates new queries combining evidence pool's content with the question for iterative retrieval. Compared to previous methods, RFM-RAG enables persistent knowledge retention and extracts critical information to retrieve.} 
\label{case2}
\end{figure}

\begin{figure*}[t!]
\centering
\includegraphics{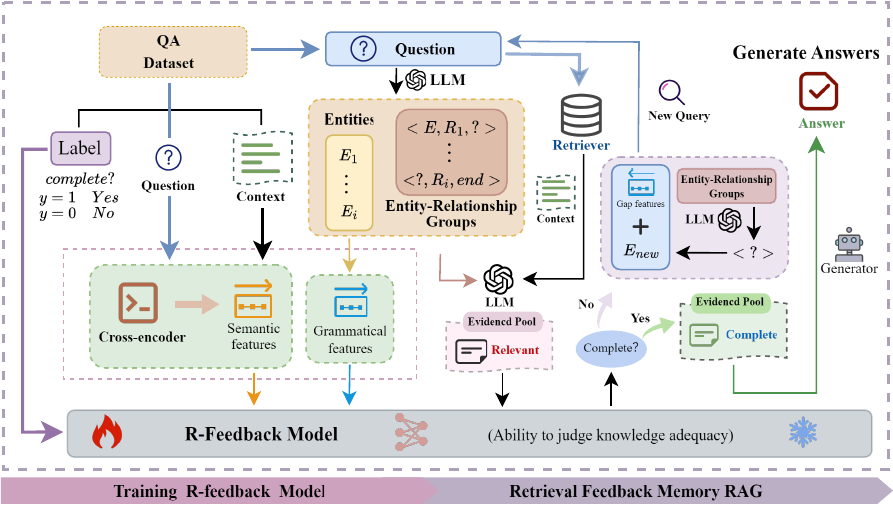}
\caption{ RFM-RAG dynamically constructs an evidence pool by processing retrieved results and formulates targeted queries until termination. The workflow begins with the original question as the initial query. An LLM then curates retrieved passages, filtering noise while integrating relevant knowledge into the evidence pool. Then, the R-Feedback Model evaluates knowledge sufficiency. If deficient, new queries are created from core question entities and evidence-pool information, iteratively enriching the evidence pool through retrieval. Upon achieving comprehensive evidence coverage, the LLM generates the final response.}
\label{overview}
\end{figure*}

In recent years, large language models (LLMs) have been widely applied in various natural language processing (NLP) tasks owing to their advanced comprehension and generation capabilities \cite{radford2018improving,chowdhery2023palm,touvron2023llama}. However, the parameter knowledge of the model remains static after pre-training. Therefore, when answering questions beyond their pretraining scope or requiring up-to-date domain knowledge, they may generate text that is syntactically fluent but factually ungrounded. This phenomenon is called hallucination \cite{maynez2020faithfulness,zhou2020detecting}.

To mitigate hallucination issues, Retrieval-Augmented Generation (RAG)\cite{lewis2020retrieval} retrieves relevant knowledge from external sources in a single pass based on user input and integrating the information into LLM prompts to enhance factual accuracy in responses. While effective for simple knowledge-intensive tasks \cite{ram2023context}, this approach performs poorly in complex scenarios requiring such as multi-step reasoning \cite{paranjape2023art}, fact verification \cite{thorne2018fever}, and Long-form generation \cite{fabbri2021summeval}.
Compared to simple tasks, these tasks require higher standards for the knowledge to be acquired. For example, Long-form generation necessitates iterative knowledge gathering throughout the generation process. Multi-hop QA requires step dependent queries where each retrieval relies on prior outputs. In contrast, iterative retrieval methods generate multiple retrieval queries and modify the retrieval queries based on feedback information through multi-round of retrieval refinement to obtain the final results\cite{asai2024self}. Dynamic retrieval RAG performs multiple retrievals during the LLM generation process. FLARE \cite{jiang2023active} uses the part of the last generated sentence to perform retrieval when the LLM’s confidence (i.e., the generation probability) on the next token is lower than certain thresholds. Some methods decompose the original question into multiple sub-questions when dealing with multi-step QA problems, retrieve external information separately and integrate multiple pieces of information as the answer.

However, iterative RAG approaches suffer from two critical limitations. Generated outputs depends on retrieved documents. low-quality retrievals introduce noise that reduce the accuracy of response. Repeated calls of the full retrieval-generation pipeline results in unnecessary resource overhead. We believe that when provided with sufficient knowledge, LLMs can generate accurate answers in a single pass. Redundant generation steps in conventional iterative RAG may have hallucinations. Thus, comprehensive knowledge completeness assessment before LLMs input is essential. Furthermore, since knowledge gaps change dynamically during iteration, each retrieval requires precise queries targeting the model's current state.

To address these limitations, we propose Retrieval-Feedback Augmented Memory Enhanced Large Model Retrieval and Generation(RFM-RAG). As shown in \textbf{Fig.\ref{case2}}, our method constructs a dynamic evidence pool where LLMs COT prompting \cite{wei2022chain} organize and deduplicate retrieved 
contexts to eliminate low-quality results. By formulating new retrieval queries through combined integration of the evidence pool and original question, we precisely target knowledge gaps beyond existing evidence. This evidence pool undergoes iterative refinement through successive queries until the Retrieval Feedback Model (R-Feedback Model) considers the evidence collection final. In summary, our main contributions are as follows:

\begin{itemize}
\item We propose an iterative retrieval-based dynamic evidence pool construction method, leveraging chain-of-thought prompts to guide LLMs in relevance filtering, structural organization, and deduplication of retrieved context. The refined information serves as validated evidence, progressively building a high-quality knowledge reservoir for final generation.
\item We design a targeted query generation mechanism that pinpoints the knowledge gaps of LLMs, enabling precise localization of missing information beyond the evidence pool. Additionally, we innovatively introduce a dedicated R-Feedback Model to evaluate the sufficiency of evidence, and release a specialized dataset for training.
\item We conduct comprehensive evaluations of previous RAG methods and RFM-RAG across three benchmark datasets using two distinct LLMs. The experimental results demonstrate improvements achieved by RFM-RAG, confirming the efficacy of our approach.
\end{itemize}

\section{Related Work}
\subsection{Retrieval-Augmented Generation}

RAG effectively mitigates hallucination with single-round retrieval enhancement being the most straightforward approach, retrieving knowledge using the original query, integrating relevant passages and prompting the LLM with augmented input.Foundational studies have extensively explored this paradigm \cite{khandelwal2019generalization, borgeaud2022improving, izacard2020leveraging, guu2020retrieval}. However, these methods are only suitable for simple tasks or unambiguous queries. 

In complex scenarios requiring multi-hop reasoning or inference, single-round retrieval often fails to capture the knowledge necessary for accurate generation precisely. As a result, recent research has focused on advanced RAG strategies. IRCot\cite{trivedi2022interleaving} employs chain-of-thought reasoning to iteratively generate retrieval queries. Adaptive-RAG\cite{jiang2023active} categorizes questions into three modes based on complexity and dynamically adjusts retrieval rounds. Self-RAG\cite{asai2024self} produces reflective tokens to guide retrieval-generation interplay. DRAGIN\cite{su2024dragin} performs real-time retrieval activated by LLM uncertainty signals during generation.

\subsection{Retrieval Quality Assessment Metrics}

Evaluating the generated outputs of large language models (LLMs) is a critical step in assessing RAG effectiveness. This process quantifies generation quality using multidimensional metrics including (factual accuracy, answer relevance, and text diversity), which collectively reflect RAG's comprehensive performance \cite{es2024ragas}. For the core retrieval component of RAG, accurate evaluation can effectively avoid unnecessary retrieval steps. Current mainstream approaches rely on quantitative metrics, which compute statistical similarity between retrieved passages and queries. Methods such as BLEU\cite{papineni2002bleu}, ROUGE\cite{lin2004rouge}, and METEOR\cite{banerjee2005meteor} evaluate relevance through surface term matching (e.g., n-gram overlap) but fundamentally ignore semantic depth. While providing measurable evaluation standards, these approaches face significant limitations in real-world applications due to insufficient understanding of deep semantics.

\section{Methodology}
Previous RAG methods suffer from over-reliance on single-round retrieval results and an inherent inability to accurately identify model knowledge gaps, frequently leading to outputs that are factually incorrect. To address these limitations, we introduce the Retrieval-Feedback Memory-enhanced RAG (RFM-RAG) framework, detailed in this section with architectural overview in \textbf{Fig.\ref{overview}}. Our methodology is based on three core principles: Dynamically constructing an evidence pool by aggregating and organizing retrieved passages for each retrieval. Using a retrieval feedback model to terminate retrieval loops upon verifying evidence pool sufficiency. Generating iterative queries through relational chain-based knowledge gap detection to address missing information.

\subsection{Dynamic Evidence Pool Construction}

\begin{figure}[htbp]
\centering
\includegraphics[scale=0.88]{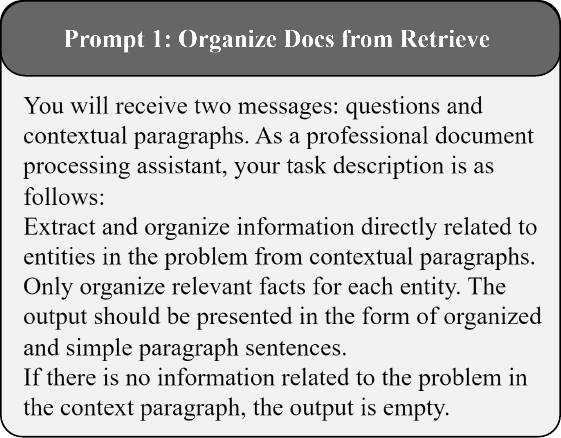}
\caption{Prompt1 for Organizing Passages Using LLMs.}
\label{prompt2}
\end{figure}

We define the vanilla LLM generation process as $\mathrm{Ans} = \mathrm{LLM}(q)$, where the LLM directly generates answers from queries. Traditional RAG methods follow $\mathrm{Ans} = \mathrm{LLM}(q, e)$ with $e = \mathcal{R}(C \mid q)$, where $\mathcal{R}$ denotes the retriever, $e$ represents relevant knowledge retrieved from corpus $C$ given $q$, and both $q$ and $e$ are input to the LLM. This paradigm suffers from incomplete retrieval and inaccurate knowledge gap identification due to the limitations of single-round retrieval. To overcome this, RFM-RAG constructs a dynamic evidence pool through iterative retrieval, leveraging R-Feedback Model(As details in the next section) to score evidence completeness and determine termination. The process initializes with the original question $q_0$. Subsequent retrievals use generated queries $q_i$, each of which obtains retrieved passages $K_i = \{k_1, k_2, \dots\}$. Using chain-of-thought prompting \cite{wei2022chain}, we instruct the LLM (GPT-3.5-turbo)\cite{brown2020language} to curate retrieved passages (As shown in \textbf{Fig.\ref{prompt2}}). This curation process involves filtering redundancies while extracting question-relevant evidence and incrementally augment the evidence pool. This evidence accumulation process is formally defined as:
\begin{equation}  
K_i = \mathcal{R}(C \mid q_i)\quad E_i = LLM_{prompt1}(q_i, K_i)
\end{equation}
\begin{equation}  
E = \{E_0, E_1, … ,E_i\}
\end{equation}

Before the R-Feedback Model decides to terminate the evidence pool construction, each retrieval iteration requires formulating a new query for extracting necessary information from external databases. Most RAG methods leverage query expansion or rewriting techniques. These methods parse semantic features and metadata within queries. However, they fail to capture critical information from retrieved passages. Consequently, we propose a query generation strategy based on knowledge gap detection, designed to more identify missing knowledge in LLM responses by detecting entity deficiencies in the query and newly emerged relevant entities in the evidence pool.

We quantify entity gaps with entity coverage metrics. Specifically, chain-of-thought prompting instructs the LLM to extract key entities $k$ and relational triples $r_k$ from the question, replacing unknown information with placeholders \textless X\textgreater. For the question \textit{Is the director of Move (1970 Film) and the director of Méditerranée (1963 Film) the same country?}, this extracts entities \textit{Move, Méditerranée} and triples \textit{(Move,director,\textless X\textgreater),(Méditerranée, director,\textless X\textgreater), (\textless X\textgreater,country,end)}. The entity coverage feature $S_{f_k}$ is computed as the proportion of knowledge about entity $k$ present in the current evidence pool $E$. When $S_{f_k}$ falls below preset threshold $\theta$, it indicates insufficient entity information in $E$, prompting addition to the gap list $G^{\prime}$ to represent unretrieved entity knowledge.
\begin{equation} 
(k, r_k)=LLM_{prompt2}(q_0)
\end{equation}
\begin{equation}
S_{f_k}= min( \frac{ C_{k_E}}{L_E}, 1.0)
\label{k_coverage}
\end{equation}
\begin{equation}
G^{\prime} = \begin{cases}
\text{Add}(G, k) & \text{if}\quad S_{f_k} < \theta \\
G & \text{otherwise}
\end{cases}
\end{equation}

where $C_{k_E}$ is the number of occurrences of entity $k$ in the current evidence pool $E$, $L_E$ is the length of the evidence pool information. Appendix C introduces the prompt templates used to extract entities and Entity-Relationship groups from the question.

\begin{figure}[htbp]
\centering
\includegraphics[scale=0.88]{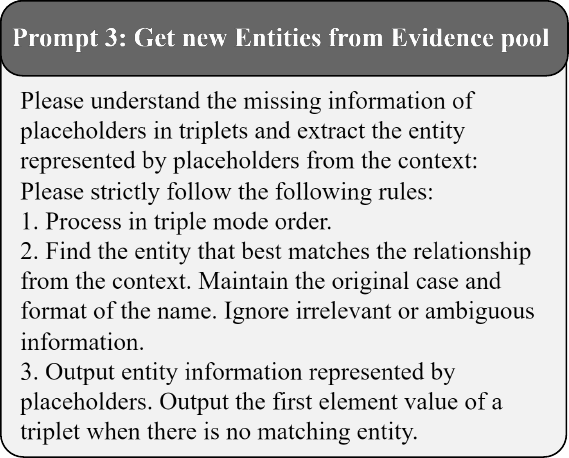}
\caption{Prompt3 for Extracting Placeholder-Represented Entities from Evidence Pool.}
\label{prompt3}
\end{figure}

Subsequently, we extract new question-related entities from the evidence pool. We input the extracted relational triples and evidence pool into the large model (As shown in \textbf{Fig.\ref{prompt3}}), enabling the model to retrieve missing information $z_k$ represented by placeholders in the triples from the evidence pool and add to the knowledge gap list $G =\{ g_1, g_2, ..., g_n, z_1, z_2, …, z_k\}$. This augmentation captures entities indirectly related to the question that cannot be obtained from the initial question. These entities form the core for subsequent retrievals. Finally, a new query $q_i$ is constructed using lexical items in $G$, designed to cover the knowledge gaps requiring external knowledge base retrieval for accurate question answering by the LLM.
\begin{equation} 
z_k=LLM_{prompt3}(r_k, E)
\end{equation}
\subsection{R-Feedback Model}
We design the R-Feedback Model as a feed-forward network with double hidden layers and activation functions. We compute the syntactic entity coverage feature $S_f$ and semantic relevance feature $G_f$ from the evidence pool $E$ and the initial question $q_0$. These features serve as input to the R-Feedback Model, which decides when to terminate evidence pool updates.

Therefore, we utilize the entity coverage calculated in Equation \ref{k_coverage} for each key entity $k$. We take the average of the coverage features of all entities to obtain the overall syntactic coverage of the entire question in the evidence pool, which is used to describe the syntactic relevance between the question and the evidence pool. Where $|E|$ is the number of entities extracted from $q_0$:
\begin{equation}  
S_f = \frac{sum(S_{f_k})}{|E|}
\end{equation}

Cross-encoders process queries and paragraphs concurrently through deep attention mechanisms, capturing complex semantic relationships with high accuracy. We derive semantic relevance features $G_f$ by processing the evidence pool $E$ and initial query $q_0$ using cross-encoder, and then fuse the two features as input to the R-Feedback Model $RF_{model}$:
\begin{equation}  
G_f =Encoder_{cross}(q_0,E)
\end{equation}
\begin{equation}  
Logit_SG = RF_{model}(S_f, G_f)
\end{equation}

Using the value of $Logit_SG$, R-Feedback Model decides if the condition for updating the evidence pool has been met.

\begin{table}[htbp]
\centering
\begin{tabular}{l|l}
\hline
\begin{tabular}[c]{@{}l@{}}\textbf{Question:} Which film \\ was released more \\ recently, \textit{Die schöne} \\ \textit{ Lurette} or \textit{Sabhash}?\end{tabular}   

& \begin{tabular}[c]{@{}l@{}}\textbf{Context:} Hobby won the \\Award for Best First Time\\ Director. Karl Geary...\\  \textbf{Label:0}\end{tabular}                                      \\ \hline

\begin{tabular}[c]{@{}l@{}}\textbf{Question:} Were \textbf{Dan}\\ \textbf{O'Connor} and \textbf{Hale} \\ \textbf{Baugh} from the same \\ country?\end{tabular}                            & 

\begin{tabular}[c]{@{}l@{}}\textbf{Context:} \textbf{Daniel O’Connor} \\was a \textbf{Canadian} politician, \\businessman... \textbf{Hale Baugh} \\ was an \textbf{American} modern \\pentathlete. He competed...\\ \textbf{Label: 1}\end{tabular}                 \\ \hline

\begin{tabular}[c]{@{}l@{}}\textbf{Question:} Which film \\ has the director who was\\ born later, \textit{Il Diavolo In } \\ \textit{Convento} or \textbf{The}\\ \textbf{Enchanting Enemy}?\end{tabular} & 

\begin{tabular}[c]{@{}l@{}}\textbf{Context:} Il diavolo in corpo\\is an... \textbf{The Enchanting} \\ \textbf{Enemy} is an Italian comedy \\film \textbf{directed by Claudio}\\ \textbf{Gora} and starring... \\ \textbf{Label: 0}\end{tabular} \\ \hline
\end{tabular}
\caption{Examples of datasets for R-Feedback Model. Information relevant to the question and context is marked in bold, and entities with missing information in the context are marked in italics.}
\label{data_case}
\end{table}

\begin{table*}[t!]
\centering
\begin{tabular}{c|c|cc|cc|c|cc}
\hline
                                     &                        & \multicolumn{2}{c|}{\textbf{2WikiMultihopQA}}      & \multicolumn{2}{c|}{\textbf{NaturalQA}}            & \textbf{StrategyQA} & \multicolumn{2}{c}{\textbf{Average}}               \\ \hline
\textbf{LLM}                         & \textbf{RAG Method}    & \multicolumn{1}{c|}{\textbf{EM}}   & \textbf{ACC}  & \multicolumn{1}{c|}{\textbf{EM}}   & \textbf{ACC}  & \textbf{ACC}        & \multicolumn{1}{c|}{\textbf{EM}}   & \textbf{ACC}  \\ \hline
\multirow{6}{*}{\textbf{Gemma-2b}}   & No Retrieval           & \multicolumn{1}{c|}{22.6}          & {\underline{43.0}}    & \multicolumn{1}{c|}{15.0}          & 24.6          & 56.0                  & \multicolumn{1}{c|}{18.8}          & 41.2          \\ \cline{2-9} 
                                     & Vanilla RAG            & \multicolumn{1}{c|}{22.8}          & 38.4          & \multicolumn{1}{c|}{11.4}          & 26.0          & 56.3                & \multicolumn{1}{c|}{17.1}          & 40.2          \\ \cline{2-9} 
                                     & Probing-RAG            & \multicolumn{1}{c|}{24.2}          & \textbf{43.6} & \multicolumn{1}{c|}{\underline{21.6}}    & {\textbf{35.0}}    & 61.8                & \multicolumn{1}{c|}{\underline{22.9}}          & \textbf{46.8}          \\ \cline{2-9} 
                                     & Adaptive RAG           & \multicolumn{1}{c|}{21.6}          & 40.6          & \multicolumn{1}{c|}{11.4}          & 26.2          & 54.7                & \multicolumn{1}{c|}{16.5}          & 40.5          \\ \cline{2-9} 
                                     & DRAGIN                 & \multicolumn{1}{c|}{\underline{26.4}}    & 28.8          & \multicolumn{1}{c|}{18.8}          & 22.2          & {\underline{62.4}}          & \multicolumn{1}{c|}{22.6}          & 37.8          \\ \cline{2-9} 
                                     & \textbf{RFM-RAG(Ours)} & \multicolumn{1}{c|}{\textbf{29.2}} & 37.6          & \multicolumn{1}{c|}{\textbf{30.6}} & \underline{33.2} & \textbf{63.2}       & \multicolumn{1}{c|}{\textbf{29.9}} & {\underline{44.7}} \\ \hline  \hline
\multirow{6}{*}{\textbf{Mistral-7b}} & No Retrieval           & \multicolumn{1}{c|}{16.4}          & 30.0          & \multicolumn{1}{c|}{13.2}          & 19.8          & 62.4                & \multicolumn{1}{c|}{14.8}          & 37.4          \\ \cline{2-9} 
                                     & Vanilla RAG            & \multicolumn{1}{c|}{21.6}          & 32.6          & \multicolumn{1}{c|}{16.8}          & 35.0          & 60.7                & \multicolumn{1}{c|}{19.2}          & 42.7          \\ \cline{2-9} 
                                     & Probing-RAG            & \multicolumn{1}{c|}{23.0}          & {\underline{33.4}}    & \multicolumn{1}{c|}{\underline{20.8}}    & {\underline{39.4}}    & 61.5                & \multicolumn{1}{c|}{\underline{21.9}}    & 44.7          \\ \cline{2-9} 
                                     & Adaptive RAG           & \multicolumn{1}{c|}{22.6}          & 31.6          & \multicolumn{1}{c|}{17.2}          & 37.4          & 65.4                & \multicolumn{1}{c|}{19.9}          & {\underline{44.8}}    \\ \cline{2-9} 
                                     & DRAGIN                 & \multicolumn{1}{c|}{\underline{23.2}}    & 25.8          & \multicolumn{1}{c|}{16.8}          & 37.2          & \underline{70.3}               & \multicolumn{1}{c|}{20.0}          & 44.4          \\ \cline{2-9} 
                                     & \textbf{RFM-RAG(Ours)} & \multicolumn{1}{c|}{\textbf{32.1}} & \textbf{36.7} & \multicolumn{1}{c|}{\textbf{33.4}} & \textbf{42.8} & \textbf{72.6}       & \multicolumn{1}{c|}{\textbf{32.7}} & \textbf{50.7} \\ \hline
\end{tabular}
\caption{ Experimental results on three different QA datasets. We indicate the highest performance in bold and underline the second highest.}
\label{results}
\end{table*}
\subsection{Training R-Feedback Model}

Training the retrieval feedback model requires dataset pairs $((q,E),y)_{1}^{N}$, where $q$ denotes the question, $E$ represents a knowledge segment, and $y \in {0,1}$ indicates sufficiency of $E$ to answer $q$. To generate these pairs, we use the evidential chain corresponding to the answer to question $q$ in the dataset to divide the context into supporting evidence and irrelevant information. Sufficient samples ($y=1$) use gold supporting evidence from the dataset as $E$, indicating $E$ fully answers $q$ without further retrieval. Insufficient samples ($y=0$) assign irrelevant information to $E$, denoting $E$ cannot answer $q$. Partially sufficient samples ($y=0$) combine subsets of supporting with irrelevant information as $E$, simulating scenarios where $E$ contains relevant but incomplete knowledge requiring additional retrieval.

As detailed in Table \ref{data_case}, our training dataset comprises three data categories derived from the public 2WikiMultihopQA \cite{ho2020constructing} dataset. To ensure a balanced distribution of positive and negative samples, we randomly selected questions and generated paired samples for each category: sufficient evidence ($y=1$) and insufficient evidence ($y=0$). The final dataset contains 10,000 training and 800 validation samples. We trained the R-Feedback Model using this dataset, with cross-entropy loss defined as follows:
$$L=-\frac{1}{N}\sum_{i=1}^N\left[y_i\log(p_i)+(1-y_i)\log(1-p_i)\right]$$

We provide details on the hyperparameters for training the R-Feedback Model in Appendix A.

\section{Experimental Setups}
\subsection{Datasets}
For performance assessment, we evaluate methods using three open-domain QA datasets, randomly sampling 500 test examples per dataset. Comprehensive dataset and corpus specifications are provided in Appendix B.

\textbf{2WikimultihopQA} \cite{ho2020constructing}. It contains multi-hop questions that span more than two Wikipedia pages, each provided with 10 paragraphs. The dataset features fine-grained paragraph annotations and a high proportion of distractors, which enables rigorous testing of models' multi-hop reasoning in noisy environments.

\textbf{NaturalQA} \cite{kwiatkowski2019natural}. Answers must be found in long documents to locate the exact fragments. Due to the real distribution of user questions and the challenge of locating answers, this task effectively tests the model's ability to extract accurate information from long, open-domain texts.

\textbf{StrategyQA} \cite{geva2021did}. It contains binary questions requiring implicit reasoning strategies without providing explicit evidence paragraphs. Characterized by strategic reasoning requirements, it assesses models' ability to construct evidence chains and perform complex inference.

\subsection{Baselines}

We choose the following Text Generation baselines for comparison. \textbf{No Retrieval}. Directly generates answers from the original question without retrieval.\textbf{Vanilla RAG}\cite{lewis2020retrieval}. Relevant passages are retrieved from an external corpus based on the initial question. The retrieved passages are then added into the LLM’s input. \textbf{DRAGIN}\cite{su2024dragin}. Retrieves when token-level confidence drops, using attention weights to construct queries from contextually salient words. \textbf{Adaptive-RAG}\cite{jeong2024adaptive}. Classifies question complexity via fine-tuned classifier to dynamically adjust retrieval steps. \textbf{Probing-RAG}\cite{baek2024probing}. Leverages intermediate-layer hidden states to determine need for additional retrieval.

All methods were evaluated under few-shot settings: 4-shot on 2WikiMultihopQA and NaturalQA, 6-shot on StrategyQA. Answer extraction used regular expression pattern matching to structure free-form LLM outputs into precise final answers. For evaluation, we used answer-level exact match(EM) and accuracy(ACC) scores to compare extracted answers against reference labels. Given diminishing accuracy gains and significant latency increases beyond three retrieval rounds, we capped maximum number of retrievals at three.

\begin{figure*}[]
\centering
\includegraphics[scale=0.42]{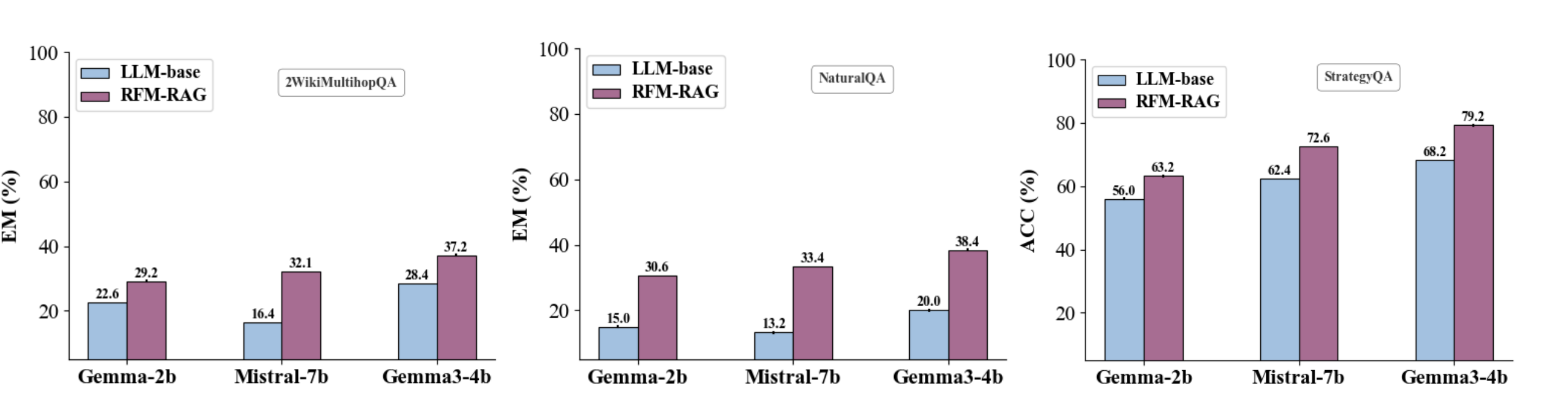}
\caption{EM and ACC scores for QA without retrieval and RFM-RAG based on Gemma-2b, Mistral-7b, and Gemma3-4b models. RFM-RAG outperforms the generation models themselves on all three datasets and all models.}
\label{exp1}
\end{figure*}

\begin{table*}[htbp]
\centering
\small
\begin{tabular}{cc|ccc|ccc}
\hline
                                                          & \multicolumn{1}{l|}{}                 & \multicolumn{3}{c|}{\textbf{NaturalQA}}          & \multicolumn{3}{c}{\textbf{StrategyQA}}          \\ \hline
                                            \multicolumn{1}{c|}{}                & \multicolumn{1}{l|}{\textbf{Methods}} & \textbf{R-Step} & \textbf{$\Delta$}          & \textbf{EM}   & \textbf{R-Step} & \textbf{$\Delta$}          & \textbf{ACC}  \\ \hline
\multicolumn{1}{c|}{\multirow{2}{*}{\textbf{Gemma-2b}}}   & \textbf{RFM}                          & 1.93             &               & \textbf{30.6} & 2.31             &               & \textbf{63.2} \\
\multicolumn{1}{c|}{}                                     & \textbf{wo-RFM}                       & 3                & \textbf{1.07$\downarrow$} & 29.5          & 3                & \textbf{0.69$\downarrow$} & 62.8          \\ \hline \hline
\multicolumn{1}{c|}{\multirow{2}{*}{\textbf{Mistral-7b}}} & \textbf{RFM}                          & 2.34             &               & 33.4          & 2.62             &               & \textbf{72.6}          \\
\multicolumn{1}{c|}{}                                     & \textbf{wo-RFM}                       & 3                & \textbf{0.66$\downarrow$} & \textbf{35.8} & 3                & \textbf{0.38$\downarrow$} &   70.2            \\ \hline
\end{tabular}

\caption{Comparison of averaged retrieval steps and EM, ACC (\%) between RFM-RAG and the ablated wo-RFM approach (Evidence pool construction termination fixed at maximum retrieval count 3) using Gemma-2b and Mistral-7b models.}
\label{exp2}
\end{table*}

\subsection{Implementation Details}

We employ BM25\cite{robertson1976relevance}, a probabilistic sparse retrieval model based on \cite{robertson2009probabilistic}, which demonstrates superior performance in RAG, even surpassing certain dense retrievers. This is implemented via ElasticSearch for all methods  to ensure fairness. For 2WikiMultihopQA, we adopt IRCoT's\cite{trivedi2022interleaving} document corpus. StrategyQA averages 2.7 evidence documents per question and has no official corpus, while NaturalQA provides only answer-containing documents. Consequently, we constructed dedicated corpora using dataset contexts (details in Appendix B). All RAG methods utilize Gemma-2b\cite{team2024gemma} and Mistral-7b\cite{Albert2023Mistral} as QA models. For computing resources, we utilize A100 GPUs with 40GB memory. In addition, due to the significant costs associated with evaluating retrieval-augmented generation models, we conducted experiments with a single run.

\section{Experimental Results}
\subsection{Main Results}

Our experiments comprehensively evaluated the performance of RFM-RAG on three datasets against various baselines, with results shown in Table \ref{results}. Our observations indicate that in most cases, single-round retrieval RAG consistently outperformed direct LLMs generation in question answering and confirming the efficacy of retrieval augmentation for knowledge-intensive QA tasks. The RFM-RAG method showed excellent performance on the majority of LLMs and datasets. Compared to no retrieval and single-round retrieval methods, on the Gemma-2b model, EM improved by approximately 11.1 and 12.8 percentage points, ACC improved by 3.5 and 4.5 percentage points. On the Mistral-7b model, EM improved by 17.9 and 13.5 percentage points, ACC improved by approximately 13.3 and 8 percentage points. This demonstrates the robustness and effectiveness of RFM-RAG in terms of knowledge collection and organization, as well as its ability to detect knowledge gaps in the model.

Notably, RFM-RAG demonstrates consistent performance gains on Gemma-2b, proving that models with fewer parameters can achieve competitive QA performance when provided with sufficient relevant information. Adaptive-RAG underperforms significantly across datasets. While it adjusts retrieval based on question complexity, the method lacks iterative enhancement targeting model’s specific knowledge gaps. The RFM-RAG we propose outperforms all previous adaptive retrieval methods by avoiding redundant generation cycles. By constructing a dynamic evidence pool through detecting model knowledge gaps, our method achieves significant performance improvements.

\begin{table*}[htbp]
\resizebox{\linewidth}{!}{%
\begin{tabular}{lll}
\hline
\textbf{Question}                                                                                 & \textbf{DRAGIN}                  & \textbf{RFM-RAG(Ours)}                                                                                                                                                                              \\ \hline
\begin{tabular}[c]{@{}l@{}}Who is the mother of \\ the director of film\\ Polish-Russian War?\\ (2WikiMultihopQA)\end{tabular} & 

\begin{tabular}[c]{@{}l@{}}\textbf{Query 1:} mother director film Polish-Russian \\War? director film Polish-Russian War\\

\textbf{Knowledge for LLM:} bombs would be like \\the early…The event attracted an audience…\\ university leaders to combat a wide array… 

\\ \textbf{Query 2:} mother director film Polish-Russian \\War… mother Andrzej Wajda Zofia Wajda.\\

\textbf{Knowledge for LLM:} bombs would be like the \\
early…The event attracted…mentioned a few \\times in the Torah and references…

\\ \textbf{Answer:} Zofia Wajda
\\ \textbf{EM: 0}\end{tabular} 
& \begin{tabular}[c]{@{}l@{}}\textbf{Query 1:} Who is the mother of the director \\of film Polish-Russian War?\\ 

\textbf{Knowledge for LLM:} The director of the film \\"Polish-Russian War" is Xawery Żuławski.\\ 
\textbf{Query 2:} Xawery Żuławski.\\ 
\textbf{Knowledge for LLM:} The director of the film \\"Polish-Russian War" is Xawery Żuławski. \\ Małgorzata Braunek is the mother of \\Xawery Żuławski, the Polish film director.\\ 
\textbf{Answer:} Małgorzata Braunek\\ 
\textbf{EM: 1}\end{tabular} \\ \hline

\begin{tabular}[c]{@{}l@{}}
what is the name of \\the rca victor dog?\\ (NaturalQA)
\end{tabular}
&  
\begin{tabular}[c]{@{}l@{}}
\textbf{Query 1:} name rca victor dog Bristol, served \\model painting Francis Barraud titled Master's.\\

\textbf{Knowledge for LLM:} were the first in provincial\\ ...Ginsberg, In this mode perfection is basic,...\\Nipper(1884–1895)was a dog from Bristol, who \\served as the model for a painting.

\\ \textbf{Query 2:} image basis dog-and-gramophone \\trademark, Berliner's successor Co. Victor Records)\\

\textbf{Knowledge for LLM:} 
were the first in provincial...\\This image was the basis for the dog-and... Berliner' \\American successor the Victor Talking Machine\\ Co. (later known as RCA Victor).\\ 
\textbf{Answer:} Berliner\\ 
\textbf{EM: 0}

\end{tabular}
&  \begin{tabular}[c]{@{}l@{}}
\textbf{Query 1:} what is the name of the rca victor dog?\\ 

\textbf{Knowledge for LLM:} Berliner's successor \\the Victor Talking Machine Co. (later known as \\RCA Victor)
\\ 

\textbf{Query 2:} Berliner\\ 
\textbf{Knowledge for LLM:} Nipper(1884–1895)was a dog, \\who served as the model for a painting.This image\\ was the basis for the dog-and-gramophone \\trademark that was used by Berliner's successor\\ the Victor Talking Machine Co.(later known as \\RCA Victor). 
\\ 
\textbf{Answer:} Nipper\\ 
\textbf{EM: 1}
\end{tabular}
\\ \hline
\end{tabular}%
} 
\caption{Case study with the RFM-RAG and DRAGIN.}

\label{case}
\end{table*}

\subsection{Analysis}

\textbf{RFM-RAG performance is unaffected by the generation model.} To investigate the impact of the generation model's inherent capabilities on the retrieval augmentation methods, we conducted supplementary experiments on the latest model, \textbf{Gemma3-4b} \cite{team2025gemma}. The experimental setup is identical to that of the main experiment: the retriever uses BM25 (implemented in ElasticSearch), the corpus is the same as in Appendix B. The prompt engineering uses the same mind chain template. \textbf{Fig.\ref{exp1}} compares the ability of the three generative models based solely on parameterized knowledge with the ability of our RFM-RAG to answer questions on three datasets. For all three models, RFM-RAG outperforms the others across all datasets. Especially for the latest generative model(Gemma3-4b), RFM-RAG improves the EM or ACC score by 8.8 points on 2WikiMultihopQA, 18.4 points on NaturalQA, and 11 points on StrategyQA, relative to the model's inherent generative capability.

\textbf{Evaluating Retrieval Feedback Model's Iteration Termination Efficacy.} Compared to fixed-iteration baselines that terminate without considering knowledge sufficiency, our method employs early termination when sufficient evidence is acquired. This strategy significantly reduces latency and mitigates noise from redundant retrievals. We empirically compared the step counts and Exact Match (EM) scores between the Fixed-iteration baseline (wo-RFM) and RFM-RAG’s adaptive termination on NaturalQA and StrategyQA datasets. Table \ref{exp2} shows that unnecessary retrieval beyond knowledge saturation leads to a reduction in accuracy by 0.4 to 2.4 percentage points on average, while RFM-RAG achieves latency reductions of 12-35\% and maintains comparable or superior accuracy, validating the efficacy of our retrieval feedback mechanism.

\textbf{Case Study.} We conducted a case study comparing RFM-RAG and DRAGIN qualitatively on 2WikiMultihopQA and NaturalQA question pairs(Table \ref{case}), analyzing retrieval queries, knowledge provisioning, and final answers. In Case 1 (complex multi-hop QA), RFM-RAG extracts key entities from retrieval results as subsequent queries. The second retrieval provides targeted knowledge for accurate answer generation. Conversely, DRAGIN relies on generation-based knowledge inference after first retrieval, introducing uncertainty. DRAGIN extracts missing knowledge from model-generated information after the initial retrieval. However, due to the uncertainty of model generation, its accuracy is weaker than the knowledge extracted from authentic evidence pool related to the question.

In Case 2, which requires information integration from multiple knowledge sources, RFM-RAG processes and retains all retrieved evidence throughout iterations. During final generation, the LLM filters relevant information from the complete evidence pool to formulate answers. DRAGIN fails to retain previously retrieved passages in subsequent retrievals. As a result, even when partial answers are generated from prior knowledge, the lack of critical evidence undermines the integrity of the final conclusion.

\section{Conclusion}

In this work, we introduce RFM-RAG, a novel retrieval pipeline that employs a relationship chain-based query generation pattern that enables precise multi-round of retrieval. During this process, the LLM organizes and deduplicates the retrieved results to construct a comprehensive evidence pool. To optimize the retrieval process, RFM-RAG incorporates an R-Feedback Model, which is responsible for determining when to stop updating the evidence pool during the retrieval rounds. This model ensures that retrievals continue only as long as necessary to gather relevant evidence. We introduce both the training dataset and method for the R-Feedback Model and show that RFM-RAG outperforms previous methods for various QA datasets.

\bibliography{aaai2026}

\end{document}